# Derivation of the Schrödinger equation from QED

Spyros Efthimiades

*Department of Natural Sciences, Fordham University*

*113 W 60th Street, New York, NY 10023*

*sefthimiades@fordham.edu*

**ABSTRACT**

The Schrödinger equation relates the electron wavefunction and the electric potential, which are emergent physical quantities. At that emergent level, the Schrödinger equation is either postulated as a principle of quantum physics or obtained heuristically. However, the Schrödinger equation is a low energy condition we can derive from the foundations of QED. Due to the small value of the electromagnetic coupling constant, we show that, in low energy interactions, the electric potential accurately represents the contributions of the intermediate photon exchanges. Then, we see that the dominant term of the electron wavefunction is a superposition of plane (but not free) waves which, by fulfilling the total energy relations, satisfies the Schrödinger, Pauli, and Dirac equations. Furthermore, we show that what is considered the kinetic energy term of the Schrödinger equation does not represent the kinetic energy of the interacting electron. We analyze and clarify the dynamics of the Schrödinger equation.

# I. INTRODUCTION

After Louis de Broglie proposed in 1924 that electrons have wave properties, Peter Debye suggested that electron waves must satisfy some wave equation. Inspired by that remark, Erwin Schrödinger proposed in 1926 that the logarithm of the electron wavefunction is proportional to the action. By requiring that the action is stationary, he obtained the equation that carries his name [1]. However, the original and similar [2] derivations are not considered adequately justified [3].

The Schrödinger equation produced breakthrough results, e.g., the electron eigenstates in the hydrogen atom. Thus, it became the "first equation of quantum mechanics" [4] and the foundation of nonrelativistic quantum physics. Because no clear-cut derivation of the Schrödinger equation was found, textbooks postulate it as a principle of quantum mechanics or obtain it heuristically.

However, the Schrödinger equation must be a low energy condition we can derive from the fundamental QED theory. Due to the small value of the electromagnetic coupling constant, we can make two critical approximations. First, we show that the electric potential accurately represents the exchanges of low energy photons between charged particles. Second, we see that the dominant term of the electron wavefunction is a superposition of plane waves that, by fulfilling the total energy relation, also satisfies the Schrödinger equation.

Furthermore, we realize and show explicitly that the assumed kinetic energy term of the Schrödinger equation does not correspond to the kinetic energy of the interacting electron.

## II. QUANTUM INTERACTIONS AND ELECTRIC POTENTIAL

The foundation of quantum electrodynamics is the electron-photon elementary interaction vertex ($e_1 + \gamma_1 \rightarrow e_2$) that connects the electron and photon waves with their interaction, whose strength is the electron charge $e$. At least one of the three particles in the interaction vertex is intermediate, so every electromagnetic process consists of two or more vertices. The wavefunction is the superposition of all contributing processes. Therefore, the electron wavefunction consists of terms proportional to powers of the coupling constant $e^2/(4\pi) = 1/137$.

The main contributions of low energy interactions between charged particles arise from the exchanges of low energy photons. The electric (Coulomb) potential represents the dominant contributions of these exchanges. We show that by considering a low energy electron-proton scattering. We disregard spins and take the proton to be immovable at $\mathbf{r} = 0$. We calculate the lowest ($e^2$) order term of the scattering amplitude $A(\theta)$ by integrating the outgoing electron waves of all processes for which the exchange of an intermediate photon changes the incoming electron momentum $\mathbf{p}$ to $\mathbf{p}_\theta$.

The intermediate photon's contribution is inversely proportional to $q^2 = q_0^2 - \mathbf{q}^2$ (its deviation from the free photon state $q^2 = 0$). Because the proton is much more massive, the electron energy does not change, so $q_0 = 0$. Integrating over $\mathbf{q}$, we get an integral expression that describes the scattering arising from the incoming electron wave interacting with the proton's electric potential $V(r) = +e/r$.

## III. ENERGY RELATION

The wavefunction of a scattered particle consists of free particle waves, whose amplitudes we can obtain from the particle's momentary interactions. In contrast, a bound electron interacts continuously with the electric potential. Therefore, we must determine how to describe the electron's interaction with a potential that varies in space and, possibly, time.

We cannot specify a particle's momentum *and* position. We represent the particle by its momentum to integrate its wave properties because the particle wavelength depends on the momentum. Thus, the amplitude of the particle wave depends on the momentum. (There are analogous energy and time dependencies.) Because the (complex) particle wave has a constant intensity, the probability of the particle's position is specified by a superposition of particle waves. So, the wavefunction $\Psi(\mathbf{r})$ of a bound electron is a superposition of electron waves.

The electron's potential energy $PE(\mathbf{r})$ equals the electric potential at $\mathbf{r}$ times the probability that the electron is there

$$PE(\mathbf{r}) = \frac{-e^2}{r} \Psi^*(\mathbf{r})\Psi(\mathbf{r}) \qquad (1)$$

Having determined how to represent the electron potential energy, we can write down the nonrelativistic energy relation.

On the macroscopic scale, the energy of a body at every point equals its kinetic plus potential energies. Given the energy and the potential, the macroscopic energy relation specifies the magnitude of the body's momentum at each point. However, such a deterministic description on the particle scale is inapplicable.

On the quantum level, we can only equate the total energy to the total kinetic plus potential energies

$$E_{total} = KE_{total} + PE_{total} \qquad (2)$$

The total energy relation is an overall requirement, not a dynamical equation.

## IV. SCHRÖDINGER EQUATION

Quantum dynamics operates on the level of particle waves and wavefunctions. We derive the Schrödinger wavefunction equation from the total energy relation by considering an electron bound by the electric potential $V(x) = +e/x$. We eliminate time dependence and get simpler expressions containing the electron energy $E$. However, $E$ is not a parameter but a derivable quantity. Only the potential determines the interaction of a bound particle.

Representing the electron wavefunction by $\Psi(x)$, the total energy relation becomes

$$E = KE + PE \quad \rightarrow \quad \int E\Psi^*\Psi\,dx = KE + \int \frac{-e^2}{x}\Psi^*\Psi\,dx \qquad (3)$$

The lowest order (dominant) term $\psi(x)$ of the wavefunction $\Psi(x)$ that satisfies the energy relation in (3) is a superposition of plane waves. We disregard electron paths with interactions because the amplitudes of their waves have $e^2/(4\pi)$ factors. So, we have

$$\psi(x) = \frac{1}{\sqrt{2\pi\hbar}} \int a(p) e^{\frac{i}{\hbar}px}\,dp \qquad (4)$$

The above expression is a turning point in the dynamical description because we represent the electron interaction with a superposition of non-interacting plane waves. However, these are not free particle waves. For example, in the hydrogen ground eigenstate, the electron plane waves have momenta from $-\infty$ to $+\infty$, but all carry $-13.6$ eV energy.

Because $\psi(x)$ is a superposition of plane waves, it is a continuous and differentiable function, and the nonrelativistic total kinetic energy can be expressed in the following forms

$$KE = \int \frac{p^2}{2m} a^*(p)a(p)\,dp = \int \frac{+\hbar^2}{2m} \frac{\partial \psi^*}{\partial x}\frac{\partial \psi}{\partial x} dx = \int \frac{-\hbar^2}{2m} \psi^* \frac{\partial^2 \psi}{\partial x^2} dx \qquad (5)$$

The last two integrals differ by a vanishing surface term. However, their integrands have different values. We will see that the integrand of the last term does not represent the electron kinetic energy.

We use the last term of (5) to extract a wavefunction equation. So, we write the total energy relation as

$$\int E\psi^*\psi\,dx = \int \frac{-\hbar^2}{2m} \psi^* \frac{\partial^2 \psi}{\partial x^2} dx + \int \frac{-e^2}{x} \psi^*\psi\,dx$$

$$\int \psi^* \left[ E\psi - \frac{-\hbar^2}{2m}\frac{\partial^2 \psi}{\partial x^2} - \frac{-e^2}{x}\psi \right] dx = 0$$

(6)

Because $\psi(x)$ is a superposition of plane waves, the integrand in the bottom equation of (6) vanishes for all $x$. To see this, let us suppose that at some point $x_i$, the bracket of the integrand is equal to $n_i$. Then, we have

$$E\psi - \frac{-\hbar^2}{2m}\frac{\partial^2 \psi}{\partial x_i^2} - \frac{-e^2}{x_i}\psi = n_i \qquad (7)$$

However, now the amplitudes $a(p)$ and the wavefunction $\psi(x)$ would depend on $n_i$, which is unphysical. Any deviation from $n_i = 0$ would affect the wavefunction globally. For example, the mild deviation $n_i = n\psi(x_i)$ changes the electron energy from $E$ to $(E–n)$. While that energy shift introduces only a phase factor to the wavefunction, nevertheless, it denotes the effect of an arbitrary interaction. In conclusion, because $\psi(x)$ is a superposition of plane waves, $n_i \neq 0$ produces unphysical results. So, the integrand in the bottom equation (6) vanishes at each point.

We have shown then that the dominant term of the electron wavefunction satisfies the following condition, which is the Schrödinger equation

$$E\psi = \frac{-\hbar^2}{2m}\frac{\partial^2 \psi}{\partial x^2} + \frac{-e^2}{x}\psi \qquad (8)$$

Following analogous steps, we derive the time-dependent Schrödinger equation in three dimensions. In particular, the dominant part $\psi(\mathbf{r},t)$ of the wavefunction satisfies the total energy relation

$$\int \psi^* \left[ i\hbar \frac{\partial \psi}{\partial t} - \frac{-\hbar^2}{2m}\nabla^2 \psi - V(\mathbf{r},t)\psi \right] d^3\mathbf{r}\, dt = 0 \qquad (9)$$

The wavefunction's dominant term $\psi(\mathbf{r},t)$ is a superposition of plane waves. Therefore, the integrand in (9) equals zero at each space-time point. So, we obtain the space-time Schrödinger equation

$$i\hbar \frac{\partial \psi}{\partial t} = \frac{-\hbar^2}{2m}\nabla^2 \psi + V(\mathbf{r},t)\psi \tag{10}$$

The solution $\psi(\mathbf{r},t)$ is a space-time function that also contains the coupling constant of the interaction.

Likewise, we derive the Pauli equation from the total energy relation that includes the magnetic energy of the electron spin. We get the Dirac equation from the total relativistic energy relation written in linear form.

We have derived the Schrödinger equation from the total energy relation, so it has its form. However, as we have indicated in our derivation and will show in the following two sections, it is not an energy conservation condition because its middle term does not correspond to the kinetic energy of the interacting electron.

Furthermore, the Schrödinger equation yields the wavefunction in collisions. For example, in the interaction region of a low-energy electron-nucleus scattering, the initial electron energy $E_{in}$ equals the total kinetic plus potential energies

$$E_{in} = KE_{total} + PE_{total} \tag{11}$$

From the above energy relation, we get that the electron wavefunction satisfies the Schrödinger equation in which the plane waves within the interaction region become the scattered electron waves.

## V. DYNAMICS OF THE SCHRÖDINGER EQUATION

To examine the physical content of the Schrödinger equation, we multiply equation (8) by $\psi^*(x)$ and get

$$E\psi^*\psi = \frac{-\hbar^2}{2m}\psi^*\frac{\partial^2\psi}{\partial x^2} + \frac{-e^2}{x}\psi^*\psi \qquad (12)$$

The first and last terms represent the bound electron's negative energy and negative potential energy at *x*. We show below that the middle term does not represent the electron kinetic energy.

The following expressions yield the nonrelativistic kinetic energy of a *single particle wave* $\psi_p(x) \sim \exp(ipx/\hbar)$

$$KE(x) = \frac{+\hbar^2}{2m}\frac{\partial \psi_p^*}{\partial x}\frac{\partial \psi_p}{\partial x} = \frac{p^2}{2m} \quad \text{and} \quad C(x) = \frac{-\hbar^2}{2m}\psi_p^*\frac{\partial^2 \psi_p}{\partial x^2} = \frac{p^2}{2m} \qquad (13)$$

For an interacting electron, *KE*(*x*) is positive and represents the superposition of the kinetic energies, while *C*(*x*) can be positive, zero, and negative. For example, *C*(*x*) in (12) is positive for small *x*, zero at $x = e^2/|E|$, and negative for large *x*. *C*(*x*) is a balancing term determined by the potential.

Kinetic energy is complementary to potential energy in the macroscopic and the total quantum energy relations. However, it is not a part of the Schrödinger equation. The Schrödinger equation is not an energy conservation condition. If it were, the electron kinetic energy would have been infinite at the nucleus. Instead, it is finite, representing the superposition at the nucleus of the kinetic energies of the plane waves that extend throughout the interaction region.

Quantum dynamics, based on particle waves, can differ remarkably from macroscopic dynamics.

## VI. THEORY AND EXPERIMENT

We consider the electron ground eigenstate in the hydrogen atom as an example of our theoretical analysis. The dominant term of the electron eigenfunction is

$$\psi(\mathbf{r},t) = \frac{1}{(2\pi\hbar)^{4/2}} \int a(\mathbf{p},E) e^{\frac{i}{\hbar}(\mathbf{p}\cdot\mathbf{r}-Et)} d^3\mathbf{p}\,dt \tag{14}$$

As there are no boundaries, the momenta of the simultaneous non-interacting electron motions have values from $-\infty$ to $+\infty$ in every direction. The Schrödinger equation for the space part $\psi(\mathbf{r})$ of the electron eigenfunction is

$$E\psi(\mathbf{r},t) = -\frac{\hbar^2}{2m}\nabla^2\psi(\mathbf{r},t) - \frac{e^2}{r}\psi(\mathbf{r},t) \tag{15}$$

Solving (15), we get the electron ground eigenfunction

$$\psi(\mathbf{r},t) = \psi(\mathbf{r}) \times e^{-\frac{i}{\hbar}Et} = \frac{1}{\sqrt{\pi r_0^3}} e^{-\frac{r}{r_0}} \times e^{-\frac{i}{\hbar}Et} \quad \text{where} \quad r_0 = \frac{\hbar^2}{me^2} = 5.29\times 10^{-11}\,m$$

$$E = \int \frac{-\hbar^2}{2m}\psi^*\nabla^2\psi\, d^3\mathbf{r} + \int \frac{-e^2}{r}\psi^*\psi\, d^3\mathbf{r} = +\frac{e^2}{2r_0} - \frac{e^2}{r_0} = -\frac{e^2}{2r_0} = -13.6\,eV \tag{16}$$

$$a(p) = \int \psi(\mathbf{r}) e^{-\frac{i}{\hbar}\mathbf{p}\cdot\mathbf{r}} d^3\mathbf{p} \quad \rightarrow \quad a(p) = \frac{1}{\pi}\left(\frac{2r_0}{\hbar}\right)^{3/2} \frac{1}{\left[1+\left(p^2/\hbar^2\right)r_0^2\right]^2}$$

The probability that the electron has momentum $p$ is $|a(p)|^2$. We measure the electron momenta by scattering photons off hydrogen atoms. The measured probabilities agree with the derived ones shown in (16) [5-6]. Theory and experiment concur that the superposition of plane waves accurately represents the electron eigenfunction. Furthermore, we have:

$$|\psi|^2 = \frac{1}{\pi r_0^3} e^{\frac{-2r}{r_0}} \qquad C(r) = \frac{-\hbar^2}{2m} \psi^* \nabla^2 \psi = \left(\frac{e^2}{r} - \frac{e^2}{2r_0}\right)|\psi|^2$$

$$PE(r) = \frac{-e^2}{r}|\psi|^2 \qquad E(r) = C(r) + PE(r) = -\frac{e^2}{2r_0}|\psi|^2$$

$$KE(r) = \frac{+\hbar^2}{2m} \nabla\psi^* \cdot \nabla\psi = \frac{e^2}{2r_0}|\psi|^2 = +13.6\,eV|\psi|^2 \qquad (17)$$

$$E = KE_{total} + PE_{total} \rightarrow \frac{-e^2}{2r_0} = \frac{e^2}{2r_0} - \frac{e^2}{r_0} = -13.6\,eV$$

$$KE(r_0) = C(r_0) \quad E(r_0) = KE(r_0) + PE(r_0) \quad KE(r_0) = -\frac{1}{2}PE(r_0)$$

$C(r)$ is positive infinite at the origin and cancels out the negative infinite potential energy term there. $C(r)$ is positive for $r < 2r_0$, zero at $r = 2r_0$, and negative for $r > 2r_0$. $KE(r)$ is positive everywhere and represents the electron kinetic energy – the superposition of the kinetic energies of the electron plane waves.

The energy relation and the virial condition hold at $r = r_0$. Furthermore, $r_0$ (the radius of the electron's first orbit in the Bohr model) is the most probable distance to find the electron.

We also realize that the interference of short wavelength (high momentum) electron waves is constructive only near the nucleus. The longer the wavelength of the electron waves, the larger the spherical volume where their interference is constructive. Thus, the farther from the nucleus we may detect the electron, the lower the momenta we will (most probably) measure. That aspect is analogous to the radius and speed dependence in macroscopic orbits.

## VII. TRANSFORMATION OF PHYSICAL DESCRIPTION

We have derived the Schrödinger equation from the form of the particle-wave and the elementary electron-photon interaction vertex

$$e\ \acute{}\ \bar{u}(p_2) g \times e(k_1) u(p_1)\, d^{(4)}(p_2 - p_1 - k_1) \qquad (18)$$

We also used that an intermediate photon's contribution is inversely proportional to its deviation from its free state and the total energy relation.

In our derivation, we have seen how low-energy approximations transform the description of the interaction. By disregarding spins and higher-order terms, we obtained the electric potential $V = e/r$. The elementary electron-photon interaction in (18) involves a photon. However, no photons appear in the Schrödinger equation because the electric potential represents the dominant contributions of low energy photon exchanges between charged particles. Furthermore, the dominant term of the electron wavefunction is a superposition of plane waves that satisfies the Schrödinger equation.

While the elementary interaction vertex incorporates energy-momentum conservation, the Schrödinger equation is not an energy conservation equation. Nevertheless, from our derivation, we know and have seen explicitly in (17) that the solutions of the Schrödinger equation fulfill the total energy relation.

In conclusion, we have seen why and how we can accurately describe low–energy quantum interactions with continuous wavefunctions that satisfy space-time differential equations.

## VIII. DISCUSSION

The Schrödinger equation is the foundation of nonrelativistic quantum mechanics.

Most textbooks state or assume that the Schrödinger equation is a postulate of quantum physics. For example, that it "plays the role that Newton's second law of motion plays in classical mechanics" [7-9].

Other textbooks start with the classical Hamiltonian and convert it to the Schrödinger equation by formally substituting energy, kinetic energy, and potential energy with operators acting on the wavefunction [10-14]. Furthermore, several textbooks state that the Schrödinger equation "cannot be obtained from some other principles since it is a fundamental law of nature." Nevertheless, they heuristically obtain it by inserting the potential energy term into the free particle wave differential equation [15-16]. Still, there is an implicit or explicit admission that by inserting "by hand" the potential term, "we make an unjustified leap" [17]. Moreover, in some textbooks, the Schrödinger equation is obtained through plausible arguments by combining total, kinetic, and potential energy terms into a space-time energy equation [18-19].

The formal, heuristic, and plausible methods consider the Schrödinger equation as an energy conservation relation. However, we have seen that what they assume to be its "kinetic energy" term does not correspond to the kinetic energy of the interacting electron. The Schrödinger equation is not an energy conservation equation. That is also the case with the Pauli equation. Because the Dirac equation is linear, it does correspond to the relativistic energy relation.

There are also attempts to derive the Schrödinger equation from classical physics, Brownian motion, and various other approaches, but none has been generally accepted [20-32].

Richard Feynman produced a path integral derivation of the Schrödinger equation [33]. Nevertheless, he maintained that "it is not possible to derive it from anything you know" [34]. Feynman was right, but not in the way he intended. We cannot derive the Schrödinger equation as a physical law because it is an approximate condition.

The above approaches to derive the Schrödinger equation are unsuccessful because they start at the emergent physical level of potential energy and electron wavefunction.

In our approach, we considered that physical consistency requires that the Schrödinger equation must be a low energy condition we can derive from the fundamental QED theory. So, we started with the elementary electron-photon interaction. First, we showed that we could accurately describe low energy photon exchanges by electric potentials that depend smoothly on position. Then we realized that the dominant term of the electron wavefunction is a superposition of plane waves that, by fulfilling the total energy relations, also satisfies the Schrödinger, Pauli, and Dirac differential equations.

There are essential differences between the typical views about the Schrödinger equation and the perceptions gained through our approach. The Schrödinger equation has been considered a principle of quantum physics. Instead, we have shown that it is a low energy condition that we can derive from the foundations of QED. Furthermore, the Schrödinger equation is considered an energy conservation relation. However, we have shown that the Schrödinger equation does not have a kinetic energy term. Because quantum dynamics is based on particle waves, it differs from macroscopic dynamics. For example, we have seen that the infinite negative potential at the atomic nucleus does not produce an infinite kinetic energy there.

The approach we followed in this paper and the results and insights we have obtained make nonrelativistic quantum dynamics clear, tangible, and physically justified.

**REFERENCES**

[1] E. Schrödinger, *Collected Papers on Wave Mechanics*, 2nd ed., Chelsea Publishing Company, New York (1978).

[2] L.D. Landau and E. M. Lifshitz, *Quantum Mechanics: Non-Relativistic Theory*, Pergamon, New York (1977).

[3] I. Bialynicki-Birula, M. Cieplak, J. Kaminski, *Theory of Quanta*, Oxford University Press (1992).

[4] R. Feynman, *The Feynman Lectures on Physics*, Addison-Wesley, Reading, Massachusetts (1965).

[5] A. Rae, *Quantum Mechanics*, 4th edition, Institute of Physics Publishing, Bristol (2002).

[6] J.W.M. Dumond and H.A. Kirkpatrick, Physical Review, **52** (1937).

[7]  D. Griffiths and D. Schroeter, *Introduction to Quantum Mechanics*, 3rd ed., Cambridge University Press (2018).

[8]  A.C. Phillips, *Introduction to Quantum Mechanics*, Wiley, Chichester (2003).

[9]  S.M. Blinder, *Introduction to Quantum Mechanics*, Elsevier Academic Press, Amsterdam (2004).

[10]  F. Dyson, *Advanced Quantum Mechanics*, World Scientific Publishing, Singapore (2007).

[11]  J.J. Sakurai, Modern Quantum Mechanics, Cummings Publishing Co., Menlo Park, Ca. (1985).

[12]  N. Zettili, *Quantum Mechanics*, Wiley, Chichester (2001).

[13]  H. Hameka, Quantum Mechanics A Conceptual Approach, Wiley, Hoboken, N.J. (2004).

[14]  L. Susskind and A. Friedman, *Quantum Mechanics, The Theoretical Minimum*, Basic Books, New York (2014).

[15]  S. Weinberg, *Foundations of Modern Physics*, Cambridge Univ. Press, Cambridge (2021).

[16]  S. Gasiorowicz, *Quantum Physics*, 2nd ed., Wiley, New York (1996).

[17]  R. Scherrer, *Quantum Mechanics: An Accessible Introduction*, Pearson, San Francisco (2006).

[18]  E. Merzbacher, *Quantum Mechanics*, 3rd ed., Willey, New York (1998).

[19]  J. Schwichtenberg, *No-Nonsense Quantum Mechanics*, No-Nonsense Books (2020).

[20] E. Nelson, "Derivation of the Schrödinger equation from Newtonian mechanics," Phys. Rev. 150, 1079-1085 (1966).

[21] E. Santamato, "Geometric derivation of the Schrödinger equation from classical mechanics in curved Weyl spaces," Phys. Rev. D 29, 216-222 (1984).

[22] B. Roy Frieden, "Fisher information as the basis for the Schrödinger wave equation," Am. J. Phys. 57, 1004-1008 (1989).

[23] C. G. Gray, G. Karl, and V. A. Novikov, "From Maupertius to Schrödinger. Quantization of classical variational principles," Am. J. Phys. 67, 959-961 (1999).

[24] M. J. W. Hall and M. Reginatto, "Schrödinger equation from an exact uncertainty principle," J. Phys. A: Math. Gen. 35, 3289-3303 (2002).

[25] L. Fritsche and M. Haugk, "A new look at the derivation of the Schrödinger equation from Newtonian mechanics," Ann. Phys. 12, 371-402 (2003).

[26] G. Grössing, "From classical Hamiltonian flow to quantum theory: derivation of the Schrödinger equation," Found. Phys. Lett. 17, 343-362 (2004).

[27] J. S Briggs, S. Boonchui, and S. Khemmani, "The derivation of time-dependent Schrödinger equations," J. Phys. A: Math. Theor. 40, 1289-1302 (2007).

[28] G. Grössing, "The vacuum fluctuation theorem: Exact Schrödinger equation via nonequilibrium thermodynamics," Phys. Lett. A 372, 4556-4563 (2008).

[29] P. R. Sarma, "Direct derivation of Schrödinger equation from Hamilton-Jacobi equation using uncertainty principle," Rom. J. Phys. 56, 1053-1056 (2011).

[30] A. Deriglazov and B. F. Rizzuti, "Reparametrization-invariant formulation of classical mechanics and the Schrödinger equation," Am J Phys 79, 882-885 (2011).

[31] J. H. Field, "Derivation of the Schrödinger equation from the Hamilton-Jacobi equation in Feynman's path integral formulation of quantum mechanics," Eur. J. Phys. 32, 63-87 (2011).

[32] G. González, "Relation between Poisson and Schrödinger equations," Am J Phys 82, 715- 719 (2012).

[33] D. Derbes, "Feynman's derivation of the Schrödinger equation," Am. J. Phys. 64, 881-884 (1996).

[34] T. Hey and P. Walters, *The New Quantum Universe*, Cambridge University Press (2009).